\documentclass[smallextended]{svjour3}       
\usepackage[english]{babel}
\usepackage{xcolor}
\usepackage{amsmath}
\usepackage{amsfonts}
\usepackage{amssymb}
\usepackage{calrsfs}
\usepackage{graphicx}
\usepackage{bussproofs}
\usepackage{proof}
\usepackage{latexsym}
\usepackage[ND,SEQ]{prftree}
\usepackage{multicol, caption}
\usepackage{setspace}
\usepackage[nottoc,numbib]{tocbibind}

\smartqed  
\usepackage{graphicx}
\usepackage{hyperref}

\begin{document}

\title{What is the meaning of proofs?}
\subtitle{A Fregean distinction in proof-theoretic semantics}

\author{Sara Ayhan\thanks{I would like to thank several people for supporting me in improving this paper essentially, among them Luca Tranchini for his thorough feedback and vital input on an earlier version of this paper and also two anonymous referees for their very constructive and helpful reports. I am especially grateful to Heinrich Wansing for the numerous and encouraging occasions to discuss this paper extensively and for his valuable comments.}         
}

\institute{Sara Ayhan \at
              Institute of Philosophy I, Ruhr University Bochum, Bochum, Germany \\
              \email{sara.ayhan@rub.de}           
}

\date{Received: date / Accepted: date}

\maketitle

This is a post-peer-review, pre-copyedit version of an article published in the Journal of Philosophical Logic.
The final authenticated version will be available online at: DOI: 10.1007/s10992-020-09577-2
\\

\begin{abstract}
The origins of proof-theoretic semantics lie in the question of what constitutes the meaning of the logical connectives and its response: the rules of inference that govern the use of the connective.
However, what if we go a step further and ask about the meaning of a proof as a whole?
In this paper we address this question and lay out a framework to distinguish sense and denotation of proofs.
Two questions are central here.
First of all, if we have two (syntactically) different derivations, does this always lead to a difference, firstly, in sense, and secondly, in denotation?
The other question is about the relation between different kinds of proof systems (here: natural deduction vs. sequent calculi) with respect to this distinction. 
Do the different forms of representing a proof necessarily correspond to a difference in how the inferential steps are given?
In our framework it will be possible to identify denotation as well as sense of proofs not only within one proof system but also between different kinds of proof systems.
Thus, we give an account to distinguish a mere syntactic divergence from a divergence in meaning and a divergence in meaning from a divergence of proof objects analogous to Frege's distinction for singular terms and sentences.

\end{abstract}

\section{Introduction}
In proof-theoretic semantics (PTS) the meaning of the logical constants is taken to be given by the rules of inference that govern their use.
As a proof is constituted by applications of rules of inference, it seems reasonable to ask what the meaning of proofs as a whole would consist of on this account.
What we are particularly interested in is a Fregean distinction between sense and denotation in the context of proofs.\footnote{We assume at least a basic familiarity with this idea, laid out in Frege's famous paper ``\"Uber Sinn und Bedeutung'', cf. \cite{Frege1} for an English translation.}
This account builds up on \cite{Tranchini2016}, where such a distinction is proposed and used in a proof-theoretic explanation of paradoxes.

The notion of denotation is nothing new in the context of proofs.
It is common in the literature on proof theory and PTS (e.g. \cite[p. 6]{Kreisel}, \cite{Prawitz1971}, \cite{M-L}) to distinguish between \textit{derivations}, as linguistic objects, and \textit{proofs}, as abstract (in the intuitionistic tradition: mental) entities.
Proofs are then said to be \textit{represented} or \textit{denoted} by derivations, i.e. the abstract proof object is the denotation of a derivation.
The notion of sense, on the other hand, has been more or less neglected.
Tranchini \cite{Tranchini2016}, therefore, made a proposal that for a derivation to have sense means to be made up of applications of correct inference rules.
While this is an interesting approach to consider, Tranchini only determines whether a proof has sense or not but does not go further into what the sense of a proof exactly consists of, so there might be further questions worth pursuing.
We will spell out an account of a distinction between sense and denotation of proofs, which can be considered a full-fledged analogy to Frege's distinction concerning singular terms and sentences.\footnote{There is some literature also in the field of proof theory concerned with this Fregean distinction, however, to our knowledge, apart from \cite{Tranchini2016} this is not concerned with the sense of derivations but with the sense of sentences: cf. P. Martin-L\"of (2001). The Sense/Reference Distinction in Constructive Semantics. Transcription of a lecture given at a conference on Frege organised by G. Sundholm at Leiden, 25 August 2001, transcription by B. Jespersen, 9 August 2002: \nolinkurl{https://www.academia.edu/25695205/The_Sense_Reference_Distinction_in_Constructive_Semantics}, or \cite{Sundholm}.}
Another question concerns the relation of different kinds of proof systems (intuitionistic natural deduction (ND) and sequent calculus (SC) systems will be considered) with respect to such a distinction.
If we have two syntactically different derivations with the same denotation in different proof systems, do they always also differ in sense or can sense be shared over different systems?
\section{Connecting structure and meaning}
The basic point of departure is the simple observation that there can be different ways leading from the same premises to the same conclusion, either in different proof systems or also within one system.
The focus in this matter so far has been on normal vs. non-normal derivations in ND and correspondingly on derivations containing cut vs. cut-free derivations in SC.
However, there can also simply be a change of the order of rule applications that can lead to syntactically different derivations from the same premises to the same conclusion.
Does this lead to a different denotation or should we say that it is only the sense that differs in such cases, while the underlying proof stays the same?
\subsection{Normal form and the denotation of derivations}
One and the same proof may be linguistically represented by different derivations. 
We will follow the general opinion in taking proofs to be the denotation - the semantic value - of (valid) derivations.
In ND a derivation in \textit{normal form} is the most direct form of representation of its denotation, i.e. the represented proof object.
For our purposes we will consider a derivation to be in normal form iff neither $\beta$- nor $\eta$-conversions (cf. rules below) can be applied to it.
A derivation in normal form in ND corresponds to a derivation in cut-free form in SC. 
In intuitionistic logic derivations in non-normal form in ND (resp. with cut in SC) can be reduced to ones in normal form (resp. cut-free form).
These are then thought to represent the same underlying proof, just one more indirectly than the other, because, as Prawitz \cite[p. 257f.]{Prawitz1971} says, they represent the same \textit{idea} this proof is based on.
In order to make sense and denotation transparent, our approach will be to encode the derivations with $\lambda$-terms.
As is well known, by the Curry-Howard-isomorphism there is a correspondence between the intuitionistic ND calculus and the simply typed $\lambda$-calculus and we can formulate the following ND-rules annotated with $\lambda$-terms together with the usual $\beta$- and $\eta$-conversions for the terms. 
The $\beta$-conversions correspond to the well-known \textit{reduction} procedures, which can be formulated for every connective in ND \cite[p. 36f.]{Prawitz1965}, while the $\eta$-conversions are usually taken to correspond to proof \textit{expansions} \cite[p. 101]{M-L}.
We use $p$, $q$, $r$,... for arbitrary atomic formulas, $A$, $B$, $C$,... for arbitrary formulas, and $\Gamma$, $\Delta$,... for sets of formulas.
$\Gamma$, $A$ stands for $\Gamma \cup \left\{A\right\}$.
For variables in terms $x$, $y$, $z$,... is used and $r$, $s$, $t$,... for arbitrary terms.
\vspace{0.5cm}

\textbf{Term-annotated ND-rules}:

\vspace{0.7cm}

\quad \quad 
\infer[\scriptstyle\supset\mathrm{I}]{\lambda x.t:A \supset B}{\infer*{t:B}{\Gamma,[x:A]}}
\quad  \quad \quad \quad
\infer[\scriptstyle\supset\mathrm{E}]{App(s, t):B}{\infer*{s: A \supset B}{{\Gamma}}  \quad \infer*{t:A}{{\Delta}}}

\vspace{0.7cm}

\quad \quad 
\infer[\scriptstyle\wedge\mathrm{I}]{\left\langle s, t\right\rangle : A \wedge B}{{\infer*{s:A}{\Gamma}} \quad {\infer*{t:B}{\Delta}}}
\quad  \quad \quad \quad
\infer[\scriptstyle\wedge\mathrm{E_{1}}]{fst(t):A}{\infer*{t:A \wedge B}{{\Gamma}}}
\quad  \quad \quad \quad
\infer[\scriptstyle\wedge\mathrm{E_{2}}]{snd(t):B}{\infer*{t:A \wedge B}{{\Gamma}}}

\vspace{0.7cm}

\quad  
\infer[\scriptstyle\vee\mathrm{I_{1}}]{\texttt{inl}s:A \vee B}{\infer*{s:A}{\Gamma}}
\quad   
\infer[\scriptstyle\vee\mathrm{I_{2}}]{\texttt{inr}s:A \vee B}{\infer*{s:B}{\Gamma}}
\quad   
\infer[\scriptstyle\vee\mathrm{E}]{\texttt{case}~r~\{x.s ~\vert~y.t\}:C}{\quad \infer*{r: A \vee B}{{\Gamma}}  \quad {\infer*{s:C}{{\Delta, [x:A]}} \quad {\infer*{t: C}{{\Theta, [y:B]}}}}}

\vspace{0.4cm}

\begin{displaymath}  
\infer[\scriptstyle\bot\mathrm{E}]{abort(t):A}{\infer*{t:\bot}{\Gamma}}
\end{displaymath}

\textbf{$\beta$-conversions:}

\begin{displaymath}
App(\lambda x.t, s) 
\rightsquigarrow
t[s/x]
\end{displaymath}

\vspace{-0.6cm}
\begin{multicols}{2}
\begin{displaymath}
fst(\left\langle s, t \right\rangle)
\rightsquigarrow
s\end{displaymath}

\begin{displaymath}
snd(\left\langle s, t \right\rangle)
\rightsquigarrow
t
\end{displaymath}
\end{multicols}

\vspace{-0.6cm}
\begin{multicols}{2}
\begin{displaymath}
\texttt{case inl}r~\{x.s ~\vert~y.t\}
\rightsquigarrow
s[r/x]\end{displaymath}

\begin{displaymath}
\texttt{case inr}r~\{x.s ~\vert~y.t\}
\rightsquigarrow
t[r/y]
\end{displaymath}
\end{multicols}

\textbf{$\eta$-conversions:}
\vspace{-0.3cm}

\begin{center}
$\lambda x.App(t, x) \rightsquigarrow t $ (if $x$ not free in $t$)
\end{center}
\vspace{-0.3cm}

\begin{displaymath}
\left\langle fst(t), snd(t) \right\rangle
\rightsquigarrow
t\end{displaymath}

\vspace{-0.3cm}
\begin{displaymath}
\texttt{case }r~\{t.\texttt{inl}t ~\vert~s.\texttt{inr}s\}
\rightsquigarrow
r\end{displaymath}

We read $x : A$ as ``$x$ is a proof of $A$".
$t[t'/x]$ means that in term $t$ every free occurrence of $x$ is substituted with $t'$.
The usual capture-avoiding requirements for variable substitution are to be observed and $\alpha$-equivalence of terms is assumed.
A term that cannot be converted by either $\beta$- or $\eta$-conversion is in normal form.

Since there is a correspondence between intuitionistic SC and intuitionistic ND, for every derivation in ND there must be a derivation in SC named by the same $\lambda$-term.
This correspondence is of course not one-to-one, but many-to-one, i.e. for each proof in ND there are at least potentially different derivations in SC.\footnote{On the complications of such a correspondence and also on giving a term-annotated version of SC cf. e.g. \cite{Prawitz1965}, \cite{Zucker}, \cite{Pottinger}, \cite{Herbelin}, \cite{BarendregtGhilezan}, \cite{NegrivonPlato}, \cite{Urban}. Term-annotated sequent calculi can be found i.a. in \cite{TS} or \cite{SU}, from which our presentation is only a notational variant.}
The following are our respective SC-rules, where we use the propositional fragment of an intuitionistic SC with independent contexts \cite[p. 89]{NegrivonPlato}.
The reduction procedures remain the same as above in ND; $\beta$-reduction corresponds to the procedures needed to establish cut-elimination, while $\eta$-conversion corresponds to what may be called ``identicals-elimination" \cite{Hacking} or ``identity atomization" \cite{Dosen2008}\footnote{Showing that it is possible to get rid of axiomatic sequents with complex formulas and derive them from atomic axiomatic sequents. This is also part of cut-elimination but in principle those are separate procedures \cite[p. 26]{Dosen2008}.}:

\vspace{0.5cm}
\textbf{Term-annotated G0ip}:
\vspace{0.7cm}

\textbf{Logical axiom:}
\begin{displaymath}
  \infer[\scriptstyle\mathrm{Rf}]{x : A \vdash x : A}{}
  \end{displaymath}

\textbf{Logical rules:}

\begin{displaymath}
\quad \quad 
  \infer[\scriptstyle\wedge\mathrm{R}]{\Gamma, \Delta \vdash \left\langle s, t\right\rangle : A \wedge B}{{\Gamma \vdash s: A} \quad {\Delta \vdash t: B}}
\quad \quad 
  \infer[\scriptstyle\wedge\mathrm{L}]{\Gamma, z: A \wedge B \vdash s[[fst(z)/x]snd(z)/y] : C}{\Gamma, x: A, y : B \vdash s : C}
\end{displaymath}

\begin{displaymath}
\quad \quad
\infer[\scriptstyle\vee\mathrm{R_{1}}]{\Gamma \vdash \texttt{inl}s :A \vee B}{\Gamma \vdash s:A}
\quad 
  \infer[\scriptstyle\vee\mathrm{R_{2}}]{\Gamma \vdash \texttt{inr}s:A \vee B}{\Gamma \vdash s:B}
\quad
\infer[\scriptstyle\vee\mathrm{L}]{\Gamma, \Delta, z:A \vee B \vdash \texttt{case z}~\{x.s ~\vert~y.t\} : C}{{\Gamma, x:A \vdash s:C} \quad {\Delta, y:B \vdash t:C}}
\end{displaymath}

\begin{displaymath}
 \quad \quad
 \infer[\scriptstyle\supset\mathrm{R}]{\Gamma \vdash \lambda x.t:A \supset B}{\Gamma, x:A \vdash t:B}
\quad \quad 
  \infer[\scriptstyle\supset\mathrm{L}]{\Gamma, \Delta, x:A \supset B \vdash s[App(x, t)/y]:C}{{\Gamma \vdash t: A} \quad {\Delta, y:B \vdash s:C}}
\end{displaymath}

\begin{displaymath}
\quad \quad
  \infer[\scriptstyle\bot\mathrm{L}]{x: \bot \vdash abort(x): C}{}
\end{displaymath}

\textbf{Structural rules}:

Weakening:

\begin{displaymath}
  \infer[\scriptstyle\mathrm{W}]{\Gamma, x:A \vdash t:C}{\Gamma \vdash t:C}
\end{displaymath}

Contraction:

\begin{displaymath}
  \infer[\scriptstyle\mathrm{C}]{\Gamma, x : A \vdash t[x/y] : C}{\Gamma, x : A, y : A \vdash t : C}
\end{displaymath}

The rule of cut
\begin{displaymath}
 \infer[\scriptstyle\mathrm{cut}]{\Gamma, \Delta \vdash s[t/x] : C}{{\Gamma \vdash t : D} \quad {\Delta, x : D \vdash s : C}}
\end{displaymath}

is admissible in G0ip.

In the left operational rules as well as in the weakening rule we have the case that variables occur beneath the line that are not explicitly mentioned above the line.
In these cases the variables must be either fresh or - together with the same type assignment - already occurring in the context $\Gamma$, $\Delta$, etc. 
Same variables can only (but need not) be chosen for the same type, i.e., if a new type occurs in a proof, then a fresh variable must be chosen.
If we would allow to chose the same variable for different types, i.e. for example to let $x:A$ and $x:B$ occur in the same derivation this would amount to assuming that arbitrarily different formulas have the same proof, which is not desirable.
\subsection{Identity of proofs and equivalence of derivations}
Figuring prominently in the literature on identity of proofs is a conjecture by Prawitz \cite[p. 257]{Prawitz1971} that two derivations represent the same proof iff they are equivalent.\footnote{Prawitz gives credit for this conjecture to Martin-L\"of. Cf. also Martin-L\"of \cite[p. 102]{M-L} on this issue, in his terminology ``definitional equality".}
This shifts the question of course to asking when two derivations can be considered equivalent.
Using the equational theory of the $\lambda$-calculus is one way to provide an answer here: terms on the right and the left hand side of the $\beta$- and $\eta$-conversions are considered denotationally equal \cite[p. 16]{Girard}.
Hence, two derivations can be considered equivalent iff they are $\beta$-$\eta$-equal (cf. \cite[p. 10]{Wideback}, \cite[p. 5]{Dosen2003}, \cite[p. 83ff.]{SU}).\footnote{There is some discussion about whether $\eta$-conversions are indeed identity-preserving. Martin-L\"of \cite[p. 100]{M-L} does not think so, for example. Prawitz \cite[p. 257]{Prawitz1971} is not clearly decided but writes in the context of identity of proofs it would seem ``unlikely that any interesting property of proofs is sensitive to differences created by an expansion". Wideb\"ack \cite{Wideback}, relating to results in the literature on the typed $\lambda$-calculus like \cite{Friedman} and \cite{Statman}, argues for $\beta$-$\eta$-equality to give the right account of identity of proofs and Girard \cite[p. 16]{Girard} does the same, although he mentions, too, that $\eta$-equations ``have never been given adequate status" compared to the $\beta$-equations.} 
The denotation is then seen to be referred to by the term that annotates the formula or sequent to be proven.
We will call this the `end-term' henceforth so that we can cover and compare both ND and SC at once.
So if we have two derivations with essentially different end-terms (in the sense that they are not belonging to the same equivalence class induced by $\beta$-$\eta$-conversion), we would say that they denote essentially different proofs.
On the other hand, for two ND-derivations, where one reduces to the other (or both reduce to the same), e.g. via normalization, we have corresponding $\lambda$-terms, one $\beta$-reducible to the other (or both $\beta$-reducible to the same term). 
In this case we would say that they refer to the same proof.
Prawitz \cite[p. 257]{Prawitz1971} stresses that this seems evident since two derivations reducing to identical normal derivations must be seen as equivalent.
Note that we can also have the case that two derivations of the same formula, which would look identical in a non-term-annotated version, here for example of ND, are distinguished on the grounds of our term annotation, like the following two derivations:

\begin{multicols}{2}

ND\textsubscript{1$p \supset (p \supset (p \wedge p))$}

ND\textsubscript{2$p \supset (p \supset (p \wedge p))$}
\end{multicols}
\vspace{-0.2cm}
\quad
\infer[\scriptstyle\supset\mathrm{I^{2}}]{\lambda y.\lambda x.\left\langle x, y \right\rangle: p \supset (p \supset (p \wedge p))}
{\infer[\scriptstyle\supset\mathrm{I^{1}}]{\lambda x.\left\langle x, y \right\rangle: p \supset (p \wedge p)}
{\infer[\scriptstyle\wedge\mathrm{I}]{\left\langle x, y \right\rangle: p \wedge p}{[x : p]^{1} \quad [y : p]^{2}}}}
\quad 
\infer[\scriptstyle\supset\mathrm{I^{2}}]{\lambda x.\lambda y.\left\langle x, y \right\rangle: p \supset (p \supset (p \wedge p))}
{\infer[\scriptstyle\supset\mathrm{I^{1}}]{\lambda y.\left\langle x, y \right\rangle: p \supset (p \wedge p)}
{\infer[\scriptstyle\wedge\mathrm{I}]{\left\langle x, y \right\rangle: p \wedge p}{[x : p]^{2} \quad [y : p]^{1}}}}

\vspace{0.2cm}

The reason for this is that it is possible to generalize these derivations in different directions, which is made explicit by the variables.
Hence, the first one can be generalized to a derivation of $B \supset (A \supset (A \wedge B))$, while the second one generalizes to $A \supset (B \supset (A \wedge B))$.\footnote{For a more detailed examination of generalization cf. \cite{Wideback} or \cite{Dosen2003}.}

So, encoding derivations with $\lambda$-terms seems like a suitable method to clarify the underlying structure of proofs.
There is one kind of conversion left, though, that needs consideration, namely what we will call \textit{permutative conversions}, or also $\gamma$-conversions.\footnote{It goes under various other names, as well, like \textit{permutation/permuting} conversions or \textit{commuting/commutative} conversions. Some also prefer ``reductions" but we will go with the - to us seemingly - more neutral ``conversions". The term $\gamma$-conversions appears in \cite{Lindley}. Cf. about these conversions in general e.g. \cite{Prawitz1971}: 251-259, \cite{Girard}: Ch. 10, \cite{Groote}, \cite{Francez}.}
They become relevant here because we have disjunction as part of our logical vocabulary.
Prawitz \cite{Prawitz1965} was the first to introduce these conversions.
In the conjunction-implication-fragment of intuitionistic propositional logic derivations in normal form satisfy the subformula property, i.e. in a normal derivation $\mathcal{D}$ of $A$ from $\Gamma$ each formula is either a subformula of $A$ or of some formula in $\Gamma$.
However, with the disjunction elimination rule this property is messed up, since we get to derive a formula $C$ from $A \vee B$ which is not necessarily related to $A$ or $B$.
That is why, in order to recover the subformula property, permutation conversions are introduced, which can be presented in their most general form in the following way:

\vspace{0.7cm}

\quad
\infer{D}{\infer[\scriptstyle\vee E]{C}{\;\;\; {\infer*{A \vee B}{\Gamma}} \quad {\infer*{C}{\Delta, \lbrack A\rbrack}} \quad {\infer*{C}{\Theta, \lbrack B \rbrack}}}}
\quad\quad
	$\rightsquigarrow$
\quad\quad
\infer[\scriptstyle\vee E]{D}{\;\;\;{\infer*{A \vee B}{\Gamma}} \quad {\infer{D}{\infer*{C}{\Delta, \lbrack A\rbrack}}} \quad {\infer{D}{\infer*{C}{\Theta, \lbrack B \rbrack}}}}

\vspace{0.5cm}

Whether or not these are supposed to be taken into the same league as $\beta$- and $\eta$-conversions in matters of identity preservation of proofs is an even bigger dispute than the one mentioned concerning $\eta$-conversions.
Prawitz \cite[p. 257]{Prawitz1971} says that while there can be no doubt about the `proper reductions' having no influence on the identity of the proof, ``[t]here may be some doubts concerning the permutative $\vee$E-[...]reductions in this connection" but does not go into that matter any further.
Since he needs these reductions to prove his normalization theorem, it seems that he would be inclined not to have too many doubts about identity preservation under the permutative conversions.
Girard \cite[p. 73]{Girard}, on the other hand, does not seem to be convinced, as he says - considering an example of permutation conversion - that we are forced to identify ``\textit{a priori} different deductions" in these cases.
Even though he accepts these conversions for technical reasons, he does not seem to be willing to really identify the underlying proof objects.
Restall\footnote{Restall, G. (2017). Proof Terms for Classical Derivations. Article in progress: \nolinkurl{https://consequently.org/papers/proof-terms.pdf}}, however, analyzing derivations by assigning to them what he calls ``proof terms" rather than $\lambda$-terms, considers the derivations above as merely distinct in representation but not in the underlying proof, which on his account is the same for both.
What is more, he does so not only for technical but rather philosophical reasons, since he claims the flow of information from premises to conclusion to be essentially the same.
Lindley \cite[p. 258]{Lindley} and Tranchini \cite[p. 1037f.]{Tranchini2018} both make a point about the connection between reductions and expansions (although they speak of certain kinds of ``generalized" expansions) on the one hand and (``generalized") permutative conversions on the other, claiming that performing a (generalized) expansion on the left hand side of the conversion above followed by a reduction (and possibly $\alpha$-conversion) just yields the right hand side.
To conclude, if we only consider the $\supset$-$\wedge$-fragment of intuitionistic propositional logic, $\beta$-$\eta$-equality is enough, but if we consider a richer vocabulary, it seems to us at least that there are substantial reasons to include permutative conversions in our equational theory.\footnote{The consequence for this paper would be of course to add ``$\gamma$-conversions" to the list of relevant conversions in our definitions about normal forms, identity of denotation, etc.}
We do not aim to make a final judgment on this issue here.
Rather, when we have laid out our distinction about sense and denotation of proofs below, we will consider the matter again and show why it makes no essential difference for our purposes whether we include permutative conversions or not.
\section{The sense of derivations}
Let us spell out at this point what exactly we will consider as the sense and also again the denotation of a derivation in our approach:
\begin{quote}
\textbf{Definition of denotation:}
The denotation of a derivation in a system with $\lambda$-term assignment is referred to by the end-term of the derivation.
Identity of denotation holds modulo belonging to the same equivalence class induced by the set of $\alpha$-, $\beta$- and $\eta$-conversions of $\lambda$-terms, i.e. derivations that are denoted by terms belonging to the same equivalence class induced by these conversions are identical, they refer to the same proof object.\footnote{We use the more accurate formulation of ``belonging to the same equivalence class" here instead of the formulation we used before of two terms ``having the same normal form". The reason for this is that while these two properties coincide for most standard cases, they do not necessarily concur when it comes to Lindley's ``general permutative conversions" or also to SC in general because in these cases the confluence property is not guaranteed. We want to thank one of the anonymous referees for indicating this important point.}

\end{quote}
\begin{quote}
\textbf{Definition of sense:}
The sense of a derivation in a system with $\lambda$-term assignment consists of the set\footnote{One could also consider the question whether multi-sets are an even better choice here, which would of course yield a much stronger differentiation of senses. The reason why we consider sets instead of multi-sets is that to us the distinctions brought about by multi-sets, by e.g. a variable occurrence more or less, do not seem to go hand in hand with substantial differences in how inferences are built up.} of $\lambda$-terms that occur within the derivation.
Only a derivation made up of applications of correct inference rules, i.e. rules that have reduction procedures, can have sense.
\end{quote}
\subsection{Change of sense due to reducibility}
Concerning a distinction between sense and denotation in the context of proofs, the rare cases where this is mentioned at all deal with derivations one of which is reducible to the other or with $\lambda$-terms which are $\beta$-convertible to the same term in normal form (cf. \cite[p. 14]{Girard}, \cite[p. 501]{Tranchini2016}, Restall 2017, p. 6).
Since Tranchini is the only one to spell out the part about sense in detail, we will briefly summarize his considerations.
As mentioned above, in his account, for a derivation to have sense means that it is made up of applications of correct inference rules.
The question to be asked then is of course what makes up \textit{correct} inference rules?
Tranchini's answer is that inference rules are correct if they have reduction procedures available, i.e. a procedure to eliminate any maximal formula resulting from an application of an introduction rule immediately followed by an elimination rule of the same connective.
From a PTS point of view, applying reduction procedures can be seen as a way of interpreting the derivation because it aims to bring the derivation to a normal form, i.e. the form in which the derivation represents the proof it denotes most directly \cite[p. 507]{Tranchini2016}.\footnote{Tranchini does not restrict his examination to derivations that normalize, though, but to the contrary, uses it to analyze non-normalizable derivations, like paradoxical ones.}
So the reduction procedures are the instructions telling us how to identify the denotation of the derivation, which for Tranchini means that they give rise to the sense of the derivation.
If we have two derivations denoting the same proof, for example, one in normal form and the other in a form that can be reduced to the former, we could say in Fregean terminology that they have the same denotation but differ in their sense because they denote the proof in different ways, one directly, the other indirectly.
So, we can take as an example the following two derivations, one in normal and one in non-normal form: 

ND\textsubscript{$p \supset p$}
\begin{displaymath}
\hspace{0.5cm}
  \prfinterspace=1.2em
	\prftree[r]{$\scriptstyle\supset\mathrm{I}$}
	{[x : p]}
	{\lambda x.x: p \supset p}
\end{displaymath}

ND\textsubscript{non-normal $p \supset  p$} 
\begin{displaymath}
\hspace{0.5cm}
  \prfinterspace=1.2em
	\prftree[r]{$\scriptstyle\wedge\mathrm{E}$}
	{\prftree[r]{$\scriptstyle\wedge\mathrm{I}$}
	{\prftree[r]{$\scriptstyle\supset\mathrm{I}$}
	{[x : p]}
	{\lambda x.x: p \supset p}}
	{\prftree[r]{$\scriptstyle\supset\mathrm{I}$}
	{[y : q]}
	{\lambda y.y: q \supset q}}
  {\left\langle \lambda x.x, \lambda y.y \right\rangle: (p \supset p) \wedge (q \supset q)}}
	{\boldsymbol{fst(\left\langle \lambda x.x, \lambda y.y \right\rangle)}: p \supset p}
\end{displaymath}

The latter obviously uses an unnecessary detour via the maximal formula $ (p \supset p) \wedge (q \supset q) $, which is introduced by conjunction introduction and then immediately eliminated again, thus, producing different and more complex terms than the former derivation.
The derivation can be easily reduced to the former, though, which can be also seen by $\beta$-reducing the term denoting the formula to be proven: 
\begin{displaymath}
\boldsymbol{fst(\left\langle \lambda x.x, \lambda y.y \right\rangle)}
\rightsquigarrow
{\lambda x.x}
\end{displaymath}

We can also give an example analogous to the one above, where a non-normal term (highlighted in bold) in SC is created by using the cut rule:\footnote{Note however, that the connection between the application of cut and the resulting non-normal term is necessary but not sufficient, i.e. there can be applications of cut not creating a non-normal term. A non-normal term is produced if both occurrences of the cut formula in the premises are principal.}

SC\textsubscript{$\vdash (p \wedge p) \supset  (p \vee p)$}
\begin{displaymath}
\hspace{0.5cm}
  \prfinterspace=1.2em
	\prftree[r]{$\scriptstyle\supset\mathrm{R}$}
  {\prftree[r]{$\scriptstyle\vee\mathrm{R}$}
	{\prftree[r]{$\scriptstyle\wedge\mathrm{L}$}
	{\prftree[r]{$\scriptstyle\mathrm{W}$}
	{\prftree[r]{$\scriptstyle\mathrm{Rf}$}
	{z : p \vdash z : p}}
	{z : p, x : p \vdash z : p}}
  {y : p \wedge p \vdash fst(y) : p}} 
	{y : p \wedge p \vdash \texttt{inl}fst(y) : p \vee p}}
	{\vdash \lambda y.\texttt{inl}fst(y) : (p \wedge p) \supset (p \vee p)}
\end{displaymath}

SC\textsubscript{cut$\vdash (p \wedge p) \supset  (p \vee p)$}
\begin{displaymath}
  \prfinterspace=1.2em
	\prftree[r]{$\scriptstyle\supset\mathrm{R}$}
  {\prftree[r]{$\scriptstyle\vee\mathrm{R}$}
	{\prftree[r]{$\boldsymbol{\scriptstyle\mathrm{cut}}$}
	{\prftree[r]{$\scriptstyle\mathrm{C}$}
	{\prftree[r]{$\scriptstyle\wedge\mathrm{R}$}
	{\prftree[r]{$\scriptstyle\wedge\mathrm{L}$}
	{\prftree[r]{$\scriptstyle\mathrm{W}$}
	{\prftree[r]{$\scriptstyle\mathrm{Rf}$}
	{z : p \vdash z : p}}
	{z : p, x : p \vdash z : p}}
	{y : p \wedge p \vdash fst(y) : p}}
	{\prftree[r]{$\scriptstyle\wedge\mathrm{L}$}
	{\prftree[r]{$\scriptstyle\mathrm{W}$}
	{\prftree[r]{$\scriptstyle\mathrm{Rf}$}
	{z : p \vdash z : p}}
	{x : p, z : p \vdash z : p}}
	{y : p \wedge p \vdash snd(y) : p}}
  {y : p \wedge p, y : p \wedge p \vdash \left\langle fst(y), snd(y)\right\rangle: p \wedge p}}
	{y : p \wedge p \vdash \left\langle fst(y), snd(y)\right\rangle: p \wedge p}}
	{\prftree[r]{$\scriptstyle\wedge\mathrm{L}$}
	{\prftree[r]{$\scriptstyle\mathrm{W}$}
	{\prftree[r]{$\scriptstyle\mathrm{Rf}$}
	{z : p \vdash z : p}}
	{z : p, x : p \vdash z : p}}
	{y : p \wedge p \vdash fst(y) : p}}
	{y : p \wedge p \vdash \boldsymbol{fst\left\langle fst(y), snd(y)\right\rangle}: p}} 
	{y : p \wedge p \vdash \boldsymbol{\texttt{inl}fst\left\langle fst(y), snd(y)\right\rangle} : p \vee p}}
	{\vdash \boldsymbol{\lambda y.\texttt{inl}fst\left\langle fst(y), snd(y)\right\rangle} : (p \wedge p) \supset (p \vee p)}
\end{displaymath}

\begin{displaymath}
\boldsymbol{\lambda y.\texttt{inl}fst\left\langle fst(y), snd(y)\right\rangle}
\rightsquigarrow
\lambda y.\texttt{inl}fst(y)
\end{displaymath}
In this case again the two derivations are essentially the same because the latter can be reduced to the former by eliminating the application of the cut rule.
Again, the proof object they represent is thus the same, only the way of making the inference, represented by the different terms occurring within the derivation, differs, i.e. the sense is different.
\subsection{Change of sense due to rule permutations}
So far we only considered the case in which there is an identity of denotation but a difference in sense of derivations due to one being represented by a $\lambda$-term in non-normal form reducible to one in normal form.
However, we want to show that this is not the only case where we can make such a distinction.
This is also the reason why our approach differs from Tranchini's (who works solely in an ND system) in how we grasp the notion of sense of a derivation. 
Following Tranchini, the derivation having sense at all depends on there being reduction procedures available for the rules that are applied in it.
Since we are also interested in a comparison of sense-and-denotation relations between ND and SC systems, our approach requires that there are reduction procedures available for the created \textit{terms}.
Thereby we will be able to cover both systems at once.

Encoding the proof systems with $\lambda$-terms also makes the connection between changing the order of the rule applications and the sense-and-denotation distinction transparent, which is the other case we want to cover.
In ND \textit{with} disjunction rules it is possible to have rule permutations producing derivations with end-terms identifiable by means of the permutative conversions.
In SC, however, there are more cases of rule permutations possible. 
When the left disjunction rule is involved, this also leads to different - though $\gamma$-equal - terms; with the left conjunction or implication rule the end-term remains completely unchanged.
Consider e.g. the following three derivations in SC of the same sequent $\vdash ((q \wedge r) \vee p) \supset ((p \vee q) \wedge (p \vee r))$:
\vspace{0.3cm}

SC$_{1\vdash ((q \wedge r) \vee p) \supset ((p \vee q) \wedge (p \vee r))}$
\vspace{-0.5cm}

\begin{displaymath}
\hspace{-0.5cm}
  \prfinterspace=1.2em
  \prftree[r]{$\scriptstyle\supset\mathrm{R}$}
	{\prftree[r]{$\underline{\scriptstyle\vee\mathrm{L}}$}
	{\prftree[r]{$\scriptstyle\mathrm{C}$}
	{\prftree[r]{$\underline{\scriptstyle\wedge\mathrm{R}}$}
	{\prftree[r]{$\boldsymbol{\scriptstyle\wedge\mathrm{L}}$}
	{\prftree[r]{$\scriptstyle\mathrm{W}$}
	{\prftree[r]{$\boldsymbol{\scriptstyle\vee\mathrm{R}}$}
	{\prftree[r]{$\scriptstyle\mathrm{Rf}$}
	{q \vdash q}}
	{q \vdash p \vee q}}
	{q, r \vdash p \vee q}}
	{q \wedge r \vdash p \vee q}}
	{\prftree[r]{$\boldsymbol{\scriptstyle\wedge\mathrm{L}}$}
	{\prftree[r]{$\scriptstyle\mathrm{W}$}
	{\prftree[r]{$\boldsymbol{\scriptstyle\vee\mathrm{R}}$}
	{\prftree[r]{$\scriptstyle\mathrm{Rf}$}
	{r \vdash r}}
	{r \vdash p \vee r}}
	{q, r \vdash p \vee r}}
	{q \wedge r \vdash p \vee r}}
	{q \wedge r, q \wedge r \vdash (p \vee q) \wedge (p \vee r)}}
	{q \wedge r \vdash (p \vee q) \wedge (p \vee r)}}
	{\prftree[r]{$\scriptstyle\mathrm{C}$}
	{\prftree[r]{$\underline{\scriptstyle\wedge\mathrm{R}}$}
	{\prftree[r]{$\scriptstyle\vee\mathrm{R}$}
	{\prftree[r]{$\scriptstyle\mathrm{Rf}$}
	{p \vdash p}}
	{p \vdash p \vee q}}
	{\prftree[r]{$\scriptstyle\vee\mathrm{R}$}
	{\prftree[r]{$\scriptstyle\mathrm{Rf}$}
	{p \vdash p}}
	{p \vdash p \vee r}}
	{p, p \vdash (p \vee q) \wedge (p \vee r)}}
	{p \vdash (p \vee q) \wedge (p \vee r)}}
	{(q \wedge r) \vee p \vdash (p \vee q) \wedge (p \vee r) }}
	{\vdash ((q \wedge r) \vee p) \supset ((p \vee q) \wedge (p \vee r))}
\end{displaymath}

SC$_{2\vdash ((q \wedge r) \vee p) \supset ((p \vee q) \wedge (p \vee r))}$
\vspace{-0.5cm}

\begin{displaymath}
\hspace{-0.5cm}
  \prfinterspace=1.2em
  \prftree[r]{$\scriptstyle\supset\mathrm{R}$}
	{\prftree[r]{$\scriptstyle\vee\mathrm{L}$}
	{\prftree[r]{$\scriptstyle\mathrm{C}$}
	{\prftree[r]{$\scriptstyle\wedge\mathrm{R}$}
	{\prftree[r]{$\boldsymbol{\scriptstyle\vee\mathrm{R}}$}
	{\prftree[r]{$\boldsymbol{\scriptstyle\wedge\mathrm{L}}$}
	{\prftree[r]{$\scriptstyle\mathrm{W}$}
	{\prftree[r]{$\scriptstyle\mathrm{Rf}$}
	{q \vdash q}}
	{q, r \vdash q}}
	{q \wedge r \vdash q}}
	{q \wedge r \vdash p \vee q}}
	{\prftree[r]{$\boldsymbol{\scriptstyle\vee\mathrm{R}}$}
	{\prftree[r]{$\boldsymbol{\scriptstyle\wedge\mathrm{L}}$}
	{\prftree[r]{$\scriptstyle\mathrm{W}$}
	{\prftree[r]{$\scriptstyle\mathrm{Rf}$}
	{r \vdash r}}
	{q, r \vdash r}}
	{q \wedge r \vdash r}}
	{q \wedge r \vdash p \vee r}}
	{q \wedge r, q \wedge r \vdash (p \vee q) \wedge (p \vee r)}}
	{q \wedge r \vdash (p \vee q) \wedge (p \vee r)}}
	{\prftree[r]{$\scriptstyle\mathrm{C}$}
	{\prftree[r]{$\scriptstyle\wedge\mathrm{R}$}
	{\prftree[r]{$\scriptstyle\vee\mathrm{R}$}
	{\prftree[r]{$\scriptstyle\mathrm{Rf}$}
	{p \vdash p}}
	{p \vdash p \vee q}}
	{\prftree[r]{$\scriptstyle\vee\mathrm{R}$}
	{\prftree[r]{$\scriptstyle\mathrm{Rf}$}
	{p \vdash p}}
	{p \vdash p \vee r}}
	{p, p \vdash (p \vee q) \wedge (p \vee r)}}
	{p \vdash (p \vee q) \wedge (p \vee r)}}
	{(q \wedge r) \vee p \vdash (p \vee q) \wedge (p \vee r) }}
	{\vdash ((q \wedge r) \vee p) \supset ((p \vee q) \wedge (p \vee r))}
	\end{displaymath}

SC$_{3\vdash ((q \wedge r) \vee p) \supset ((p \vee q) \wedge (p \vee r))}$
\vspace{-0.5cm}

\begin{displaymath}
\hspace{-0.5cm}
  \prfinterspace=1.2em
  \prftree[r]{$\scriptstyle\supset\mathrm{R}$}
	{\prftree[r]{$\scriptstyle\mathrm{C}$}
	{\prftree[r]{$\underline{\scriptstyle\wedge\mathrm{R}}$}
	{\prftree[r]{$\underline{\scriptstyle\vee\mathrm{L}}$}
	{\prftree[r]{$\boldsymbol{\scriptstyle\wedge\mathrm{L}}$}
	{\prftree[r]{$\scriptstyle\mathrm{W}$}
	{\prftree[r]{$\boldsymbol{\scriptstyle\vee\mathrm{R}}$}
	{\prftree[r]{$\scriptstyle\mathrm{Rf}$}
	{q \vdash q}}
	{q \vdash p \vee q}}
	{q, r \vdash p \vee q}}
	{q \wedge r \vdash p \vee q}}
	{\prftree[r]{$\scriptstyle\vee\mathrm{R}$}
	{\prftree[r]{$\scriptstyle\mathrm{Rf}$}
	{p \vdash p}}
	{p \vdash p \vee q}}
	{(q \wedge r) \vee p \vdash p \vee q}}
	{\prftree[r]{$\underline{\scriptstyle\vee\mathrm{L}}$}
	{\prftree[r]{$\boldsymbol{\scriptstyle\wedge\mathrm{L}}$}
	{\prftree[r]{$\scriptstyle\mathrm{W}$}
	{\prftree[r]{$\boldsymbol{\scriptstyle\vee\mathrm{R}}$}
	{\prftree[r]{$\scriptstyle\mathrm{Rf}$}
	{r \vdash r}}
	{r \vdash p \vee r}}
	{q, r \vdash p \vee r}}
	{q \wedge r \vdash p \vee r}}
	{\prftree[r]{$\scriptstyle\vee\mathrm{R}$}
	{\prftree[r]{$\scriptstyle\mathrm{Rf}$}
	{p \vdash p}}
	{p \vdash p \vee r}}
	{(q \wedge r) \vee p \vdash p \vee r}}
	{(q \wedge r) \vee p, (q \wedge r) \vee p \vdash (p \vee q) \wedge (p \vee r) }}
	{(q \wedge r) \vee p \vdash (p \vee q) \wedge (p \vee r) }}
	{\vdash ((q \wedge r) \vee p) \supset ((p \vee q) \wedge (p \vee r))}
	\end{displaymath}

The difference between SC\textsubscript{1} and SC\textsubscript{2} (highlighted in bold) is that the order of applying the right disjunction rule and the left conjunction rule is permuted.
The difference between SC\textsubscript{1} and SC\textsubscript{3} (highlighted with underlining) is that the order of applying the right conjunction rule and the left disjunction rule is permuted.
The order of applying the right disjunction rule and the left conjunction rule stays fixed this time.
Encoded with $\lambda$-terms, though, we see that in the first case, comparing SC\textsubscript{1} and SC\textsubscript{2}, the permutation of rule applications produces exactly the same end-term.
Both derivations have the same end-term, namely:
\begin{center}
 $\boldsymbol{\lambda u.\texttt{case u}~\{v.\left\langle \texttt{inr}fst(v), \texttt{inr}snd(v) \right\rangle ~\vert~x.\left\langle \texttt{inl}x, \texttt{inl}x\right\rangle\}}$ 
\end{center}

SC$_{1\vdash ((q \wedge r) \vee p) \supset ((p \vee q) \wedge (p \vee r))}$

{\footnotesize
\begin{displaymath}
\hspace{-0.5cm}
  \prfinterspace=1.2em
  \prftree[r]{$\scriptstyle\supset\mathrm{R}$}
	{\prftree[r]{$\scriptstyle\vee\mathrm{L}$}
	{\prftree[r]{$\scriptstyle\mathrm{C}$}
	{\prftree[r]{$\scriptstyle\wedge\mathrm{R}$}
	{\prftree[r]{$\boldsymbol{\scriptstyle\wedge\mathrm{L}}$}
	{\prftree[r]{$\scriptstyle\mathrm{W}$}
	{\prftree[r]{$\boldsymbol{\scriptstyle\vee\mathrm{R}}$}
	{\prftree[r]{$\scriptstyle\mathrm{Rf}$}
	{y : q \vdash y : q}}
	{y : q \vdash \texttt{inr}y : p \vee q}}
	{y : q, z : r \vdash \texttt{inr}y : p \vee q}}
	{v : q \wedge r \vdash \texttt{inr}fst(v) : p \vee q}}
	{\prftree[r]{$\boldsymbol{\scriptstyle\wedge\mathrm{L}}$}
	{\prftree[r]{$\scriptstyle\mathrm{W}$}
	{\prftree[r]{$\boldsymbol{\scriptstyle\vee\mathrm{R}}$}
	{\prftree[r]{$\scriptstyle\mathrm{Rf}$}
	{z : r \vdash z : r}}
	{z : r \vdash \texttt{inr}z : p \vee r}}
	{y : q, z : r \vdash \texttt{inr}z : p \vee r}}
	{v : q \wedge r \vdash \texttt{inr}snd(v): p \vee r}}
	{v : q \wedge r, v : q \wedge r \vdash \left\langle \texttt{inr}fst(v), \texttt{inr}snd(v) \right\rangle : (p \vee q) \wedge (p \vee r)}}
	{v : q \wedge r \vdash \left\langle \texttt{inr}fst(v), \texttt{inr}snd(v) \right\rangle : (p \vee q) \wedge (p \vee r)}}
	{\prftree[r]{$\scriptstyle\mathrm{C}$}
	{\prftree[r]{$\scriptstyle\wedge\mathrm{R}$}
	{\prftree[r]{$\scriptstyle\vee\mathrm{R}$}
	{\prftree[r]{$\scriptstyle\mathrm{Rf}$}
	{x : p \vdash x : p}}
	{x : p \vdash \texttt{inl}x : p \vee q}}
	{\prftree[r]{$\scriptstyle\vee\mathrm{R}$}
	{\prftree[r]{$\scriptstyle\mathrm{Rf}$}
	{x : p \vdash x : p}}
	{x : p \vdash \texttt{inl}x : p \vee r}}
	{x : p, x : p \vdash \left\langle \texttt{inl}x, \texttt{inl}x\right\rangle: (p \vee q) \wedge (p \vee r)}}
	{x : p \vdash \left\langle \texttt{inl}x, \texttt{inl}x\right\rangle: (p \vee q) \wedge (p \vee r)}}
	{u : (q \wedge r) \vee p \vdash \texttt{case u}~\{v.\left\langle \texttt{inr}fst(v), \texttt{inr}snd(v) \right\rangle ~\vert~x.\left\langle \texttt{inl}x, \texttt{inl}x\right\rangle\}  : (p \vee q) \wedge (p \vee r) }}
	{\vdash \boldsymbol{\lambda u.\texttt{case u}~\{v.\left\langle \texttt{inr}fst(v), \texttt{inr}snd(v) \right\rangle ~\vert~x.\left\langle \texttt{inl}x, \texttt{inl}x\right\rangle\}} \color{black}: ((q \wedge r) \vee p) \supset ((p \vee q) \wedge (p \vee r))}
\end{displaymath}}

SC$_{2\vdash ((q \wedge r) \vee p) \supset ((p \vee q) \wedge (p \vee r))}$

{\footnotesize
\begin{displaymath}
\hspace{-0.5cm}
  \prfinterspace=1.2em
  \prftree[r]{$\scriptstyle\supset\mathrm{R}$}
	{\prftree[r]{$\scriptstyle\vee\mathrm{L}$}
	{\prftree[r]{$\scriptstyle\mathrm{C}$}
	{\prftree[r]{$\scriptstyle\wedge\mathrm{R}$}
	{\prftree[r]{$\boldsymbol{\scriptstyle\vee\mathrm{R}}$}
	{\prftree[r]{$\boldsymbol{\scriptstyle\wedge\mathrm{L}}$}
	{\prftree[r]{$\scriptstyle\mathrm{W}$}
	{\prftree[r]{$\scriptstyle\mathrm{Rf}$}
	{y : q \vdash y : q}}
	{y : q, z : r \vdash y : q}}
	{v : q \wedge r \vdash fst(v) : q}}
	{v : q \wedge r \vdash \texttt{inr}fst(v) : p \vee q}}
	{\prftree[r]{$\boldsymbol{\scriptstyle\vee\mathrm{R}}$}
	{\prftree[r]{$\boldsymbol{\scriptstyle\wedge\mathrm{L}}$}
	{\prftree[r]{$\scriptstyle\mathrm{W}$}
	{\prftree[r]{$\scriptstyle\mathrm{Rf}$}
	{z : r \vdash z : r}}
	{y : q, z : r \vdash z : r}}
	{v : q \wedge r \vdash snd(v) : r}}
	{v : q \wedge r \vdash \texttt{inr}snd(v): p \vee r}}
	{v : q \wedge r, v : q \wedge r \vdash \left\langle \texttt{inr}fst(v), \texttt{inr}snd(v) \right\rangle : (p \vee q) \wedge (p \vee r)}}
	{v : q \wedge r \vdash \left\langle \texttt{inr}fst(v), \texttt{inr}snd(v) \right\rangle : (p \vee q) \wedge (p \vee r)}}
	{\prftree[r]{$\scriptstyle\mathrm{C}$}
	{\prftree[r]{$\scriptstyle\wedge\mathrm{R}$}
	{\prftree[r]{$\scriptstyle\vee\mathrm{R}$}
	{\prftree[r]{$\scriptstyle\mathrm{Rf}$}
	{x : p \vdash x : p}}
	{x : p \vdash \texttt{inl}x : p \vee q}}
	{\prftree[r]{$\scriptstyle\vee\mathrm{R}$}
	{\prftree[r]{$\scriptstyle\mathrm{Rf}$}
	{x : p \vdash x : p}}
	{x : p \vdash \texttt{inl}x : p \vee r}}
	{x : p, x : p \vdash \left\langle \texttt{inl}x, \texttt{inl}x\right\rangle: (p \vee q) \wedge (p \vee r)}}
	{x : p \vdash \left\langle \texttt{inl}x, \texttt{inl}x\right\rangle: (p \vee q) \wedge (p \vee r)}}
	{u : (q \wedge r) \vee p \vdash \texttt{case u}~\{v.\left\langle \texttt{inr}fst(v), \texttt{inr}snd(v) \right\rangle ~\vert~x.\left\langle \texttt{inl}x, \texttt{inl}x\right\rangle\}  : (p \vee q) \wedge (p \vee r) }}
	{\vdash \boldsymbol{\lambda u.\texttt{case u}~\{v.\left\langle \texttt{inr}fst(v), \texttt{inr}snd(v) \right\rangle ~\vert~x.\left\langle \texttt{inl}x, \texttt{inl}x\right\rangle\}} \color{black} : ((q \wedge r) \vee p) \supset ((p \vee q) \wedge (p \vee r))}
\end{displaymath}}

Considering the second comparison between SC\textsubscript{1} and SC\textsubscript{3} the situation is different: here the permutation of rule applications leads to a different end-term.
In the end-term for SC\textsubscript{1} and SC\textsubscript{2} the pairing operation is embedded within the case expression, whereas in the end-term for SC\textsubscript{3} the case expression is embedded within the pairing:
\begin{center}
$\lambda u.\left\langle \texttt{case u}~\{v.\texttt{inr}fst(v) ~\vert~x.\texttt{inl}x\}, \texttt{case u}~\{v.\texttt{inr}snd(v)~\vert~x.\texttt{inl}x\}\right\rangle$
\end{center}

SC$_{3\vdash ((q \wedge r) \vee p) \supset ((p \vee q) \wedge (p \vee r))}$

{\footnotesize
\begin{displaymath}
\hspace{-0.5cm}
  \prfinterspace=1.2em
  \prftree[r]{$\scriptstyle\supset\mathrm{R}$}
	{\prftree[r]{$\scriptstyle\mathrm{C}$}
	{\prftree[r]{$\underline{\scriptstyle\wedge\mathrm{R}}$}
	{\prftree[r]{$\underline{\scriptstyle\vee\mathrm{L}}$}
	{\prftree[r]{$\boldsymbol{\scriptstyle\wedge\mathrm{L}}$}
	{\prftree[r]{$\scriptstyle\mathrm{W}$}
	{\prftree[r]{$\boldsymbol{\scriptstyle\vee\mathrm{R}}$}
	{\prftree[r]{$\scriptstyle\mathrm{Rf}$}
	{y : q \vdash y : q}}
	{y : q \vdash \texttt{inr}y : p \vee q}}
	{y : q, z : r \vdash \texttt{inr}y : p \vee q}}
	{v : q \wedge r \vdash \texttt{inr}fst(v) : p \vee q}}
	{\prftree[r]{$\scriptstyle\vee\mathrm{R}$}
	{\prftree[r]{$\scriptstyle\mathrm{Rf}$}
	{x : p \vdash x : p}}
	{x : p \vdash \texttt{inl}x : p \vee q}}
	{u : (q \wedge r) \vee p \vdash \texttt{case u}~\{v.\texttt{inr}fst(v) ~\vert~x.\texttt{inl}x\} : p \vee q}}
	{\prftree[r]{$\underline{\scriptstyle\vee\mathrm{L}}$}
	{\prftree[r]{$\boldsymbol{\scriptstyle\wedge\mathrm{L}}$}
	{\prftree[r]{$\scriptstyle\mathrm{W}$}
	{\prftree[r]{$\boldsymbol{\scriptstyle\vee\mathrm{R}}$}
	{\prftree[r]{$\scriptstyle\mathrm{Rf}$}
	{z : r \vdash z : r}}
	{z : r \vdash \texttt{inr}z : p \vee r}}
	{y : q, z : r \vdash \texttt{inr}z : p \vee r}}
	{v : q \wedge r \vdash \texttt{inr}snd(v) : p \vee r}}
	{\prftree[r]{$\scriptstyle\vee\mathrm{R}$}
	{\prftree[r]{$\scriptstyle\mathrm{Rf}$}
	{x : p \vdash x : p}}
	{x : p \vdash \texttt{inl}x : p \vee r}}
	{u : (q \wedge r) \vee p \vdash \texttt{case u}~\{v.\texttt{inr}snd(v) ~\vert~x.\texttt{inl}x\}: p \vee r}}
	{u : (q \wedge r) \vee p, u : (q \wedge r) \vee p \vdash \left\langle \texttt{case u}~\{v.\texttt{inr}fst(v) ~\vert~x.\texttt{inl}x\}, \texttt{case u}~\{v.\texttt{inr}snd(v) ~\vert~x.\texttt{inl}x\}\right\rangle : (p \vee q) \wedge (p \vee r) }}
	{u : (q \wedge r) \vee p \vdash \left\langle \texttt{case u}~\{v.\texttt{inr}fst(v) ~\vert~x.\texttt{inl}x\}, \texttt{case u}~\{v.\texttt{inr}snd(v) ~\vert~x.\texttt{inl}x\}\right\rangle : (p \vee q) \wedge (p \vee r) }}
	{\vdash \lambda u.\left\langle \texttt{case u}~\{v.\texttt{inr}fst(v) ~\vert~x.\texttt{inl}x\}, \texttt{case u}~\{v.\texttt{inr}snd(v) ~\vert~x.\texttt{inl}x\}\right\rangle : ((q \wedge r) \vee p) \supset ((p \vee q) \wedge (p \vee r))}
\end{displaymath}}

When we take a look at how the term-annotated rules must be designed in order to have a correspondence to the respective rules in ND, we see why some permutations of rule applications lead to different end-terms, while others do not; and why SC is in general more flexible in this respect than ND.
In SC the left conjunction rule as well as the left implication rule are substitution operations, i.e. they can change their place in the order without affecting the basic term structure because only in the inner term structure terms are substituted with other terms.\footnote{For $\supset$L the only exception is when an application of this rule is permuted with an application of $\vee$L, which creates a different, though $\gamma$-convertible term.}
In ND, on the other hand, there are no substitution operations used in the term assignment, i.e. for each rule application a new basic term structure is created.

How is this related to the distinction between sense and denotation?
In cases like SC\textsubscript{1} vs. SC\textsubscript{2} the way the inference is given differs, which can also be seen in different terms annotating the formulas occurring within the derivation: with otherwise identical terms in the two derivations \texttt{\textbf{inr}}\textbf{y} and \textbf{\texttt{inr}z} only occur in SC\textsubscript{1}, while \textbf{fst(v)} and \textbf{snd(v)} only occur in SC\textsubscript{2}.
However, the resulting end-term stays the same, thus, we would describe the difference between these derivations as a difference in sense but not in denotation.
In other cases, when disjunction elimination or the left disjunction rule is involved, permutation of rule applications can lead to a different end-term, as we see above in SC\textsubscript{1} vs. SC\textsubscript{3}.
Whether this corresponds to a difference in denotation depends on whether we accept $\gamma$-conversions to be identity-preserving.
What all cases have in common, though, is that rule permutation always leads to a difference in sense of the given derivations because the sets of terms occurring within the derivations differ from each other.

\subsection{Philosophical motivation}
Let us have a look at how the Fregean conception of sense is received in the literature in order to show the philosophical motivation for adopting such a definition of sense for derivations.
According to Dummett \cite[p. 91]{Dummett}, Fregean sense is to be considered as a procedure to determine its denotation.\footnote{This idea of sense as procedures also occurs in more recent publications like \cite{Muskens} or \cite{DJM}.}
Girard \cite[p. 2]{Girard}, in a passage about sense and denotation and the relation between proofs and programs, mentions that the sense is determined by a ``sequence of instructions" and when we see in this context terms as representing programs and ``the purpose of a program [...] to calculate [...] its denotation" (ibid., p. 17), then it seems plausible to view the terms occurring within the derivation, decorating the intermediate steps in the construction of the complex end-term that decorates the conclusion, as the sense of that derivation.
Tranchini holds the reduction procedures to be the sense because these `instructions' lead to the term in normal form.
However, in our framework - because we do not only consider normal vs. non-normal cases - it seems more plausible to look at the exact terms occurring within the derivations and view them as representing the steps in the process of construction encoding how the derivation is built up and leading us to the denotation, the end-term. 
For us it is therefore only a necessary requirement for the derivation to have sense to contain only terms for which reduction procedures are available but it does not \textit{make up} the sense.
In the case of rule permutation we can then say that the proof is essentially the same but the way it is given to us, the way of inference, differs: i.e. the sense differs.
This can be read off from the set of terms that occur within the derivation: they end up building the same end-term, but the way it is built differs, the \textit{procedures to determine the denotation} differ.
Thus, this allows us to compare differences in sense within one proof system as well as over different proof systems.

Troelstra and Schwichtenberg \cite[p. 74]{TS} e.g. give an example of two derivations in SC producing the same end-term in different ways to show that just from the variables and the end-term we cannot read off how the derivation is built up:\footnote{For simplicity we omit the weakening steps that would strictly seen have to precede the applications of the $\wedge$L-rule.}
\begin{center}
SC\textsubscript{1$\vdash (s \wedge p) \supset ((q \wedge r) \supset (p \wedge q))$}
\begin{displaymath}
  \prfinterspace=1.2em
  \prftree[r]{$\scriptstyle\supset\mathrm{R}$}
	{\prftree[r]{$\scriptstyle\supset\mathrm{R}$}
	{\prftree[r]{$\scriptstyle\wedge\mathrm{L}$}
	{\prftree[r]{$\scriptstyle\wedge\mathrm{L}$}
	{\prftree[r]{$\scriptstyle\wedge\mathrm{R}$}
	{\prftree[r]{$\scriptstyle\mathrm{Rf}$}
	{x : p \vdash x : p}}
	{\prftree[r]{$\scriptstyle\mathrm{Rf}$}
	{y : q \vdash y : q}}
  {x : p, y : q \vdash \left\langle x, y \right\rangle: p \wedge q}}
	{x : p, z : q \wedge r \vdash \left\langle x, fst(z) \right\rangle: p \wedge q}}
	{u: s \wedge p, z : q \wedge r \vdash \left\langle snd(u), fst(z) \right\rangle: p \wedge q}}
	{u : s \wedge p \vdash \lambda z.\left\langle snd(u), fst(z) \right\rangle: (q \wedge r) \supset (p \wedge q)}}
	{\vdash\lambda u.\lambda z.\left\langle snd(u), fst(z) \right\rangle: (s \wedge p) \supset ((q \wedge r) \supset (p \wedge q))}
\end{displaymath}
\end{center}

\begin{center}
SC\textsubscript{2$\vdash (s \wedge p) \supset ((q \wedge r) \supset (p \wedge q))$}
\begin{displaymath}
  \prfinterspace=1.2em
  \prftree[r]{$\scriptstyle\supset\mathrm{R}$}
	{\prftree[r]{$\scriptstyle\supset\mathrm{R}$}
	{\prftree[r]{$\scriptstyle\wedge\mathrm{L}$}
	{\prftree[r]{$\scriptstyle\wedge\mathrm{L}$}
	{\prftree[r]{$\scriptstyle\wedge\mathrm{R}$}
	{\prftree[r]{$\scriptstyle\mathrm{Rf}$}
	{x : p \vdash x : p}}
	{\prftree[r]{$\scriptstyle\mathrm{Rf}$}
	{y : q \vdash y : q}}
  {x : p, y : q \vdash \left\langle x, y \right\rangle: p \wedge q}}
	{u : s \wedge p, y: q \vdash \left\langle snd(u), y \right\rangle: p \wedge q}}
	{u: s \wedge p, z : q \wedge r \vdash \left\langle snd(u), fst(z) \right\rangle: p \wedge q}}
	{u : s \wedge p \vdash \lambda z.\left\langle snd(u), fst(z) \right\rangle: (q \wedge r) \supset (p \wedge q)}}
	{\vdash\lambda u.\lambda z.\left\langle snd(u), fst(z) \right\rangle: (s \wedge p) \supset ((q \wedge r) \supset (p \wedge q))}
\end{displaymath}
\end{center}
\vspace{0.2cm}

The senses of these derivations would be the following:
\vspace{0.5cm}

	Sense of SC\textsubscript{1}: 
	\vspace{-0.3cm}
\begin{multline*}
		\{x, y, z, u, \left\langle x, y \right\rangle, \underline{\left\langle x, fst(z) \right\rangle}, \left\langle snd(u), fst(z) \right\rangle, \lambda z.\left\langle snd(u), fst(z) \right\rangle,\\
		\lambda u.\lambda z.\left\langle snd(u), fst(z) \right\rangle \}
\end{multline*}		

Sense of SC\textsubscript{2}:
	\vspace{-0.3cm}
		\begin{multline*}
		\{x, y, z, u, \left\langle x, y \right\rangle, \underline{\left\langle snd(u), y \right\rangle}, \left\langle snd(u), fst(z) \right\rangle, \lambda z.\left\langle snd(u), fst(z) \right\rangle,\\ \lambda u.\lambda z.\left\langle snd(u), fst(z) \right\rangle\}	
		\end{multline*}

The two sets only differ with regard to the underlined terms, otherwise they are identical.
Thus, they only differ in the order in which the two left conjunction rules are applied.
For the resulting end-term this is inessential, but we can see that when taking the sense, and not only the end-terms, i.e. the denotation, into account, it is indeed possible to read off the structure of the derivations.
As noted above (examples on p. 6), the term annotation of the calculi makes this structure of derivations explicit so that we can differentiate between derivations which would otherwise look identical.
As several authors point out, this is a desirable feature if one is not only interested in mere provability but wants to study the structure of the derivations in question (cf. \cite[p. 82]{SU}, \cite[p. 93]{Pfenning}) and also, for simplicity, if one wants to compare proof systems of ND and SC with each other \cite[p. 73]{TS}.
Since we are interested in both of these points, it seems the right choice for our purposes to consider the annotated versions of the calculi and that is also why these annotated versions are indeed needed for our notions of sense and denotation.
Of course, one could argue that the underlying structure is still the same in the non-annotated versions and can be made explicit by other means, too, like showing the different generalizations of the derivations, but still, we do not see how in these calculi our notions could be easily applied.

Another issue that needs to be considered is the one of \textit{identity of senses}, i.e. \textit{synonymy}.
Therefore, we want to extend our definition of sense given above with an addition:
\begin{quote} 
If a sense-representing set can be obtained from another by uniformly replacing (respecting the usual capture-avoiding conventions) any occurrence of a variable, bound or free, by another variable of the same type, they express the same sense.
\end{quote}
What we ensure with this point is just that it does not (and should not) matter which variables one chooses for which proposition as long as one does it consistently.
So, it does not make a difference whether we have 

\begin{center}
\begin{multicols}{2}
ND\textsubscript{1$p \supset (q \supset p)$}	
\begin{displaymath}
\hspace{0.5cm}
  \prfinterspace=1.2em
	\prftree[r]{$\scriptstyle\supset\mathrm{I}$}
	{\prftree[r]{$\scriptstyle\supset\mathrm{I}$}
	{[x : p]}
	{\lambda z.x: q \supset p}}
	{\lambda x. \lambda z.x: p \supset (q\supset p)}
\end{displaymath}

Sense\textsubscript{1}: \{$x, \lambda z.x, \lambda x. \lambda z.x$\}
\end{multicols}
\end{center}

or
\vspace{-0.2cm}
\begin{center}
\begin{multicols}{2}
ND\textsubscript{2$p \supset (q \supset p)$}	
\begin{displaymath}
\hspace{0.5cm}
  \prfinterspace=1.2em
	\prftree[r]{$\scriptstyle\supset\mathrm{I}$}
	{\prftree[r]{$\scriptstyle\supset\mathrm{I}$}
	{[y : p]}
	{\lambda z.y: q \supset p}}
	{\lambda y. \lambda z.y: p \supset (q\supset p)}
\end{displaymath}

Sense\textsubscript{2}: \{$y, \lambda z.y, \lambda y. \lambda z.y$\}
\end{multicols}
\end{center}

Sense\textsubscript{1} and Sense\textsubscript{2} represent the same sense.
Or to give another example (pointed to by one of the anonymous referees) where we have free variables occurring within the derivation but not appearing in the end-term: If one would replace all occurrences of the free variable $y$ by the variable $w$ in derivation SC\textsubscript{1$\vdash (s \wedge p) \supset ((q \wedge r) \supset (p \wedge q))$} (cf. above), then this would make no difference to the sense according to our definition since the sense-representing sets would be obtained from replacing $y$ by $w$.

This also fits the Fregean criterion of two sentences' identical sense, as Sundholm \cite[p. 304]{Sundholm} depicts it within a broader analysis: two propositions express the same sense if it is not possible to hold different \textit{epistemic attitudes} towards them, i.e. ``if one holds the one true, one also \textit{must} hold the other one true, and \textit{vice versa}".
Whereas, if we have two sentences which only differ in two singular terms, referring to the same object but differing in sense, we can easily hold the one sentence to be true, while thinking the other is false, if we do not know that they are referring to the same object.
With proofs it is the same: Looking at ND\textsubscript{1$p \supset (q \supset p)$} and ND\textsubscript{2$p \supset (q \supset p)$} we may not know whether the derivation is valid or not, we \textit{do} know, however, that if one is a valid derivation then so is the other.
With derivations differing in sense this is not so straightforward.

For Frege this point of considering cases where intensionality is directed towards sentences was crucial to develop his notion of sense, so the question arises how we can explain cases of intensionality directed towards \textit{proofs} with our notions of sense and denotation.
Let us suppose we have two denotationally-identical proofs which are represented by two different derivations $\mathcal{D}$ and $\mathcal{D'}$.
In this case it could happen that a (rational) person believes that derivation $\mathcal{D}$ is valid but does not believe that derivation $\mathcal{D'}$ is valid.
How can we account for that?
One explanation would be of course to point to the difference in linguistic representation.
After all, it can just be the case that one way of writing down a proof is more accessible to the person than another (they may not be familiar with a certain proof system, for example).
This would amount to letting the linguistic representation, the signs, collapse with the sense of a derivation.
However, then we would have no means to distinguish this case from cases in which we want to argue that it is not justified for a rational person to have different propositional attitudes towards propositions which are about derivations differing insignificantly from each other, like in the cases of ND\textsubscript{1$p \supset (q \supset p)$}	and ND\textsubscript{2$p \supset (q \supset p)$} above.
For Frege \cite[p. 212, 218]{Frege1} the referent of an expression in an intensional context is not its \textit{customary referent}, i.e. the object it refers to or the truth value in the case of sentences, but its \textit{customary sense}.
Here the situation is the same: What is referred to in such a setting, when speaking about the attitudes of a person towards propositions about derivations, is not the proof objects (which are identical in our situation) but their senses, which are in this context represented by the sets of terms encoding the steps of construction.
It seems plausible then to say that when the construction steps differ in two derivations, a person can have different attitudes towards propositions about them, because the different construction steps may lead to this person grasping the one derivation, while not understanding the other.

\section{Analogy to Frege's cases}
Let us finally compare how our conception of sense and denotation in the context of proofs fits the distinction Frege came up with for singular terms and sentences.
We can have the following two cases with Frege's distinction: firstly (cf. \cite[p. 211]{Frege1}), there can be different signs corresponding to exactly one sense (and then of course also only one denotation).
In the case of singular terms an example would be ``Gottlob's brother'' and ``the brother of Gottlob".
The sense, the way the denoted individual object is given to us, is the same because there is only a minor grammatical difference between the two expressions.
More frequently, this occurs in comparing different languages, though, taking singular terms which express exactly the same sense only using different words, like ``the capital of France" and ``die Hauptstadt Frankreichs". 
In the case of sentences an example would be changing from an active to a passive construction without changing the emphasis of the sentence; an example from Frege is the following: ``$M$ gave document $A$ to $N$", ``Document $A$ was given to $N$ by $M$" \cite[p. 141]{Frege2}.
In the case of proofs, finally, an example would be the following case:
\vspace{0.3cm}

ND\textsubscript{$ (p\vee p) \supset (p\wedge p)$}
\vspace{-0.3cm}

\begin{displaymath}
\hspace{0.5cm}
  \prfinterspace=1.2em
  \prftree[r]{$\scriptstyle\supset\mathrm{I^{3}}$}
	{\prftree[r]{$\scriptstyle\wedge\mathrm{I}$}
	{\prftree[r]{$\scriptstyle\vee\mathrm{E^{1}}$}
	{[y : p \vee p]^{3}}
	{[x : p]^{1}}
	{[x : p]^{1}}
  {\texttt{case y}~\{x.x~\vert~x.x\} : p}} 
	{\prftree[r]{$\scriptstyle\vee\mathrm{E^{2}}$}
	{[y : p \vee p]^{3}}
	{[x : p]^{2}}
	{[x : p]^{2}}
  {\texttt{case y}~\{x.x~\vert~x.x\} : p}}
	{\left\langle \texttt{case y}~\{x.x~\vert~x.x\}, \texttt{case y}~\{x.x~\vert~x.x\}\right\rangle : p \wedge p}}
	{\lambda y.\left\langle \texttt{case y}~\{x.x~\vert~x.x\}, \texttt{case y}~\{x.x~\vert~x.x\}\right\rangle\color{black} : (p \vee p) \supset (p \wedge p)}
\end{displaymath}

SC\textsubscript{$\vdash (p\vee p) \supset (p \wedge p)$}
\vspace{-0.3cm}

{\small
\begin{displaymath}
\hspace{0.5cm}
  \prfinterspace=1.2em
  \prftree[r]{$\scriptstyle\supset\mathrm{R}$}
	{\prftree[r]{$\scriptstyle\mathrm{C}$}
	{\prftree[r]{$\scriptstyle\wedge\mathrm{R}$}
	{\prftree[r]{$\scriptstyle\vee\mathrm{L}$}
	{\prftree[r]{$\scriptstyle\mathrm{Rf}$}
	{x : p \vdash x : p}}
	{\prftree[r]{$\scriptstyle\mathrm{Rf}$}
	{x : p \vdash x : p}}
	{y : p \vee p \vdash \texttt{case y}~\{x.x~\vert~x.x\} : p}}
	{\prftree[r]{$\scriptstyle\vee\mathrm{L}$}
	{\prftree[r]{$\scriptstyle\mathrm{Rf}$}
	{x : p \vdash x : p}}
	{\prftree[r]{$\scriptstyle\mathrm{Rf}$}
	{x : p \vdash x : p}}
	{y : p \vee p \vdash \texttt{case y}~\{x.x~\vert~x.x\}: p}}
	{y : p \vee p , y : p \vee p \vdash \left\langle \texttt{case y}~\{x.x~\vert~x.x\}, \texttt{case y}~\{x.x~\vert~x.x\} \right\rangle : p \wedge p}}
	{y : p \vee p \vdash \left\langle \texttt{case y}~\{x.x~\vert~x.x\}, \texttt{case y}~\{x.x~\vert~x.x\}\right\rangle: p \wedge p}}
	{\vdash \lambda y.\left\langle \texttt{case y}~\{x.x~\vert~x.x\}, \texttt{case y}~\{x.x~\vert~x.x\}\right\rangle \color{black}: (p \vee p) \supset (p \wedge p)}
\end{displaymath}}

Sense: 
\begin{multline*}
\{x, y, \texttt{case y}~\{x.x~\vert~x.x\}, \left\langle \texttt{case y}~\{x.x~\vert~x.x\}, \texttt{case y}~\{x.x~\vert~x.x\}\right\rangle,\\ \lambda y.\left\langle \texttt{case y}~\{x.x~\vert~x.x\}, \texttt{case y}~\{x.x~\vert~x.x\}\right\rangle\}
\end{multline*}
\\Or to give another example:
\vspace{0.3cm}

ND\textsubscript{$p \supset (p \supset (p \wedge p))$}
\vspace{-0.5cm}

\begin{displaymath}
\hspace{0.5cm}
  \prfinterspace=1.2em
  \prftree[r]{$\scriptstyle\supset\mathrm{I^{2}}$}
	{\prftree[r]{$\scriptstyle\supset\mathrm{I^{1}}$}
	{\prftree[r]{$\scriptstyle\wedge\mathrm{I}$}
	{[x : p]^{2}}
	{[y : p]^{1}}
  {\left\langle x, y \right\rangle: p \wedge p}}
	{\lambda y.\left\langle x, y \right\rangle: p \supset (p \wedge p)}}
	{\lambda x.\lambda y.\left\langle x, y \right\rangle: p \supset (p \supset (p \wedge p))}
\end{displaymath}

SC\textsubscript{$\vdash p \supset (p \supset (p \wedge p))$}
\vspace{-0.5cm}

\begin{displaymath}
\hspace{0.5cm}
  \prfinterspace=1.2em
  \prftree[r]{$\scriptstyle\supset\mathrm{R}$}
	{\prftree[r]{$\scriptstyle\supset\mathrm{R}$}
	{\prftree[r]{$\scriptstyle\wedge\mathrm{R}$}
	{\prftree[r]{$\scriptstyle\mathrm{Rf}$}
	{x : p \vdash x : p}}
	{\prftree[r]{$\scriptstyle\mathrm{Rf}$}
	{y : p \vdash y : p}}
  {x : p, y : p \vdash \left\langle x, y \right\rangle: p \wedge p}}
	{x : p \vdash \lambda y.\left\langle x, y \right\rangle: p \supset (p \wedge p)}}
	{\vdash \lambda x.\lambda y.\left\langle x, y \right\rangle: p \supset (p \supset (p \wedge p))}
\end{displaymath}

Sense: $\{x, y, \left\langle x, y \right\rangle, \lambda y.\left\langle x, y \right\rangle, \lambda x.\lambda y.\left\langle x, y \right\rangle\} $\\

In these cases derivations can consist of different signs, namely by having one representation in SC and one in ND, which do not differ in sense nor in denotation, since they both contain exactly the same terms and produce the same end-term.
This comparison between different proof systems seems to fit nicely with Frege's \cite[p. 211]{Frege1} comment on ``the same sense ha[ving] different expressions in different languages". 
However, as we have seen above with the examples ND\textsubscript{1$p \supset (q \supset p)$} and ND\textsubscript{2$p \supset (q \supset p)$}, this case can also occur within the same proof system.
One could wonder whether there should not be a differentiation between the senses of the derivations in the first example since it seems that different rules are applied: in SC\textsubscript{$\vdash (p\vee p) \supset (p \wedge p)$} we have an application of contraction, which we do not have in ND\textsubscript{$ (p\vee p) \supset (p\wedge p)$}.
This would also question whether our definition of sense distinguishes and identifies the right amount of cases.
We do believe that this is the case, though, because in the first example, where there is an application of the contraction rule in SC, there is also a multiple assumption discharge in the ND-derivation, which is generally seen as the corresponding procedure, just as cases of vacuous discharge of assumptions in ND correspond to the application of weakening in SC. 
So just as in different languages of course not exactly the same expressions are used, here too, the rules differ from ND to SC but since the \textit{corresponding} procedures are used, one can argue that the sense does not differ for that reason.
\\
\\Another case that can occur according to Frege (ibid.) is that we have one denotation, i.e. one object a sign refers to, but different senses.
An example for this would be his famous ``morning star" and ``evening star" comparison, where both expressions refer to the same object, the planet Venus, but the denoted object is given differently.
On the sentence level this would amount to exchanging singular terms in a sentence by ones which have the same denotation: ``The morning star is the planet Venus" and ``The evening star is the planet Venus".
The denotation of the sentence - with Frege: its truth value - thus stays the same, only the sense of it differs, the information is conveyed differently to us.
For our proof cases we can say that this case is given when we have syntactically different derivations, be it in one or in different proof systems, which have end-terms belonging to the same equivalence class induced by the set of $\alpha$-, $\beta$- and $\eta$-conversions.
Thus, examples would be corresponding proofs in ND and SC, which share the same end-term, but contain different terms occurring within the derivations.
The reason for this to happen seems that in SC often more variables are necessary than in ND.
If we compare derivations \textit{within ND}, one definite case in which we have the same denotation but a different sense is between equivalent but syntactically distinct derivations, e.g. non-normal and normal derivations, one reducible to the other.
Another case up for debate would be the one with rule permutations due to disjunction elimination.
\textit{Within SC} we can have two cases: one due to rule permutation, one due to applications of cut.
For the first case, where the inference could be given in a different way, although ending on the same term, we gave examples above (cf. p. 12 and 14f.).
However, it is worth mentioning that our distinction still captures the usual distinction, the second case, where it is said that two derivations, one containing cut and the other one in cut-free form (as a result of cut-elimination applied to the former), have the same denotation but differ in sense:
\vspace{0.3cm}

SC\textsubscript{$\vdash (p\wedge p) \supset (p\vee p)$}
\vspace{-0.5cm}

\begin{displaymath}
\hspace{0.5cm}
  \prfinterspace=1.2em
  \prftree[r]{$\scriptstyle\supset\mathrm{R}$}
	{\prftree[r]{$\scriptstyle\vee\mathrm{R}$}
	{\prftree[r]{$\scriptstyle\wedge\mathrm{L}$}
	{\prftree[r]{$\scriptstyle\mathrm{W}$}
	{\prftree[r]{$\scriptstyle\mathrm{Rf}$}
	{z : p \vdash z : p}}
  {z : p, x : p \vdash z : p}}
	{y : p\wedge p \vdash fst(y) : p}} 
	{y : p \wedge p \vdash \texttt{inl}fst(y) : p \vee p}}
	{\vdash \lambda y.\texttt{inl}fst(y) : (p \wedge p) \supset (p \vee p)}	
\end{displaymath}

Sense: $\{z, x, y, fst(y), \texttt{inl}fst(y), \lambda y.\texttt{inl}fst(y)\}$
\vspace{0.3cm}

SC\textsubscript{cut$\vdash (p\wedge p) \supset (p\vee p)$}
 
\begin{displaymath}
\hspace{0.5cm}
  \prfinterspace=1.2em
  \prftree[r]{$\scriptstyle\supset\mathrm{R}$}
	{\prftree[r]{$\scriptstyle\mathrm{cut}$}
	{\prftree[r]{$\scriptstyle\wedge\mathrm{L}$}
	{\prftree[r]{$\scriptstyle\mathrm{W}$}
	{\prftree[r]{$\scriptstyle\mathrm{Rf}$}
	{z : p \vdash z : p}}
	{z : p, x : p \vdash z : p}}
	{y : p \wedge p \vdash fst(y) : p}}
	{\prftree[r]{$\scriptstyle\vee\mathrm{R}$}
	{\prftree[r]{$\scriptstyle\mathrm{Rf}$}
	{z : p \vdash z : p}}
	{z : p \vdash \texttt{inl}z: p \vee p}}
	{y : p \wedge p \vdash \texttt{inl}fst(y) : p \vee p}}
	{\vdash \lambda y.\texttt{inl}fst(y): (p \wedge p) \supset (p \vee p)}
\end{displaymath}

Sense: $\{z, x, y, fst(y), \underline{\texttt{inl}z}, \texttt{inl}fst(y), \lambda y.\texttt{inl}fst(y)\}$\\

As mentioned above (fn 14), cut does not need to create a non-normal term, as it is the case here, but still any application of cut will necessarily change the sense of a derivation as opposed to its cut-free form.

Finally, cases that need to be avoided in a formal language according to Frege \cite[p. 211]{Frege1} would be to have one sign, corresponding to different senses, or on the other hand, one sense corresponding to different denotations.
As he mentions, these cases of course occur in natural languages but should not happen in formal ones, so it should also not be possible in our present context, for sure.
Fortunately, this cannot happen in the context of our annotated proof systems, either, since the signs (taken to be the derivation as it is written down) always express at most one sense in our annotated system, and likewise the sense always yields a unique denotation since the end-term is part of the sense-denoting set.\footnote{Another question would be whether there can be signs without any sense at all. Frege \cite[p. 211]{Frege1} dismisses this case, as well, with a remark that we need at least the requirement that our expressions are ``grammatically well-formed". Tranchini \cite{Tranchini2016} gives a good analogy pointing to the notorious connective \texttt{tonk} playing this role in the case of proofs.}

\section{Conclusion}
The context in which Frege considered sense and denotation was the context of identity.
Likewise, we argued in this paper, if we use term-annotated calculi, we can also say something about proof identity: identity of proofs over different calculi or within the same calculus consists in having end-terms that belong to the same equivalence class induced by the set of $\alpha$-, $\beta$- and $\eta$-conversions.
In ND this can happen when we have the same proof in normal and non-normal form, in SC this can happen when we have the same proof using cut and in cut-free form but also when there are forms of rule permutations where an application of the $\wedge$L-rule or the $\supset$L-rule switches place with another rule.
Including disjunction in our language creates for both calculi the additional question of whether rule permutations including disjunction elimination (resp. the left disjunction rule) lead to a different proof, or whether these proofs should be identified.
We are more interested in sense, however, and here we can conclude that what in all these cases changes is the sense of the derivation in question.
Finally, considering the question of identity of sense, i.e. synonymy, and trying to follow Frege's conception on this matter, too, we can say the following: if two derivations are supposed to be \textit{identical in sense}, this means that the way the inference is given is essentially the same, so the set of terms building up the end-term must be the same.
The end-term itself does not necessarily tell us anything about the structure of the proof.
Sense, on the other hand, is more fine-grained in that the set of terms occurring within the derivation reflects how the derivation is built up.
Especially in SC, where we can have different orders of rule applications leading up to the same end-term, the sense gives us means to distinguish on a more fine-grained level.\\

\end{document}